\documentclass[a4paper,11pt]{article}
\usepackage{pos}
\usepackage{amsmath,bm,longtable,mathrsfs,slashed}

\title{Axial U(1) symmetry at high temperatures in $N_f=2+1$ lattice QCD with chiral fermions}
\ShortTitle{Axial U(1) symmetry at high temperatures in $N_f=2+1$ lattice QCD with chiral fermions}

\manuallySeparateAuthors
\author[a]{Sinya Aoki}
\author[b]{, Yasumichi Aoki}
\author[c]{, Hidenori Fukaya}
\author[d,e]{, Shoji Hashimoto}
\author[b]{, \\Issaku Kanamori}
\author[d,e]{, Takashi Kaneko}
\author[b]{, Yoshifumi Nakamura}
\author[c]{, Christian Rohrhofer}
\author[]{, and}
\author*[f]{ Kei Suzuki}
\author[]{(JLQCD Collaboration)}

\affiliation[a]{Center for Gravitational Physics, Yukawa Institute for Theoretical Physics, Kyoto 606-8502, Japan}
\affiliation[b]{RIKEN Center for Computational Science, Kobe 650-0047, Japan}
\affiliation[c]{Department of Physics, Osaka University, Toyonaka 560-0043, Japan}
\affiliation[d]{KEK Theory Center, High Energy Accelerator Research Organization (KEK), Tsukuba 305-0801, Japan}
\affiliation[e]{School of High Energy Accelerator Science, The Graduate University for Advanced Studies (Sokendai), Tsukuba 305-0801, Japan}
\affiliation[f]{Advanced Science Research Center, Japan Atomic Energy Agency (JAEA), Tokai 319-1195, Japan}

\emailAdd{k.suzuki.2010@th.phys.titech.ac.jp}

\abstract{
We study the $U(1)_A$ anomaly in the high-temperature phase of $N_f=2+1$ lattice QCD with chiral fermions.
Gauge ensembles are generated with M\"obius domain-wall (MDW) fermions, and in the measurements the determinant is reweighted to that of overlap fermions.
We report the results for the overlap Dirac spectrum, $U(1)_A$ susceptibility, and topological susceptibility at $T=204$ and $175$ MeV.
}

\FullConference{%
 The 38th International Symposium on Lattice Field Theory, LATTICE2021
  26th-30th July, 2021
  Zoom/Gather@Massachusetts Institute of Technology
}

\begin{document}
\begin{flushright}
OU-HET-1137
\end{flushright}
\maketitle

\section{Introduction}\label{sec-1}

The $SU(2)_L \times SU(2)_R$ chiral symmetry (in the massless limit of the up and down quarks) and the $U(1)_A$ anomaly pray important roles in quantum chromodynamics (QCD).
At zero temperature, the spontaneous breaking of chiral symmetry, characterized by the chiral condensate, and also the $U(1)_A$ anomaly are related to low-energy dynamics of QCD, such as masses of hadrons.
The chiral symmetry is restored if the temperature is high enough, but the fate of the $U(1)_A$ anomaly in the chiral-symmetric phase is still a problem we have not reached to a consensus.

The JLQCD collaboration has been investigating the high-temperature phase of QCD with $N_f=2$ lattice simulations~\cite{Cossu:2013uua,Tomiya:2016jwr,Aoki:2020noz}.
In Ref.~\cite{Cossu:2013uua}, we focused on the behavior of the $U(1)_A$ anomaly in a fixed topological sector.
In Ref.~\cite{Tomiya:2016jwr,Aoki:2020noz}, we generated gauge ensembles with M\"obius domain-wall (MDW) fermions \cite{Brower:2004xi,Brower:2005qw,Brower:2012vk} and quantities measured on the MDW fermion ensembles are reweighted to those on overlap (OV) fermion ensembles by using the MDW/OV reweighting technique \cite{Fukaya:2013vka,Tomiya:2016jwr}.

In these $N_f=2$ lattice QCD studies with chiral fermions~\cite{Cossu:2013uua,Tomiya:2016jwr,Aoki:2020noz}, we have found that the axial $U(1)$ anomaly is strongly suppressed near
the chiral limit, at least, to a few \% level of the scale of temperature.
Our data of the axial $U(1)$ breaking suppression show a consistency among the different lattice sizes up to $L=4$fm, as well as among different lattice spacings down to $a=0.075$ fm.

In this contribution to the proceedings of LATTICE2021, we report on our preliminary investigation of the same observables but with a more realistic set-up in $N_f=2+1$ lattice QCD.
Including the dynamical strange quark, we generate the gauge link ensembles employing the tree-level Symanzik improved gauge action~\cite{Luscher:1985zq} and dynamical MDW fermion action and the configurations are then reweighted to those with overlap fermion determinant, where the reweighting factor is stochastically computed.
Numerical parameters are summarized in Table \ref{Tab:param}.
We set the gauge coupling to $\beta=4.17$, which corresponds to the lattice cut-off $a^{-1}=2.453$ GeV (or $a\sim 0.08$ fm) and choose two different temperatures by setting the temporal lattice extent to $L_t=12$ (204 MeV) and $L_t=14$ (175 MeV), which are above the (pseudo) critical temperature $\sim 150$ MeV.
The fixed strange quark mass $m_s=0.04$ and the lightest up and down quark masses $m_{ud}=0.002$ are near their physical points, and the spacial lattice size $L=32$ is $\sim 2.6$ fm.

\begin{table}[b!]
\centering
\caption{Numerical parameters of lattice simulations.
$L^3 \times L_t $ are the spatial and temporal lattice size.
$m$ is the degenerate mass of up and down quarks, and $m_s$ is the strange-quark mass.
Lattice cutoff is $a^{-1}=2.453$ GeV.
}
\small
\begin{tabular}{ccccc}
\hline\hline
$L^3 \times L_t $ &  $T$ (MeV) & $am$ & $m$ (MeV) & $m_s$ \\
\hline
$32^3 \times 12$ & 204 & 0.0020  & 4.9 & 0.040 \\
$32^3 \times 12$ & 204 & 0.0035  & 8.6 & 0.040 \\
$32^3 \times 12$ & 204 & 0.0070  & 17 & 0.040 \\
$32^3 \times 12$ & 204 & 0.0120  & 29 & 0.040 \\
\hline
$32^3 \times 14$ & 175 & 0.0020  & 4.9 & 0.040 \\
$32^3 \times 14$ & 175 & 0.0035  & 8.6 & 0.040 \\
$32^3 \times 14$ & 175 & 0.0070  & 17  & 0.040 \\
$32^3 \times 14$ & 175 & 0.0120  & 29  & 0.040 \\
\hline\hline
\end{tabular}
\label{Tab:param}
\end{table}

\section{Overlap Dirac spectrum}\label{sec-2}
First, we study the spectral density of Dirac operator,
\begin{equation}
\rho(\lambda)=\frac{1}{V} \sum_{\lambda_i} \langle \delta(\lambda-\lambda_i) \rangle,
\end{equation}
where $V$ is the lattice volume, $\lambda_i$ is the $i$-th positive
eigenvalue of the overlap Dirac operator, and $\langle \cdots \rangle$ denotes the gauge ensemble average.
According to the Banks-Casher relation~\cite{Banks:1979yr}, the Dirac spectrum $\rho(\lambda)$ at $\lambda=0$ is proportional to the chiral condensate.
Its derivatives are also connected to the observables probing the $SU(2)_L\times SU(2)_R$ symmetry breaking, as well as to those related to the $U(1)_A$ breaking~\cite{Aoki:2012yj}.

\begin{figure}[b!]
    \begin{minipage}[t]{0.5\columnwidth}
    \includegraphics[clip,width=1.0\columnwidth]{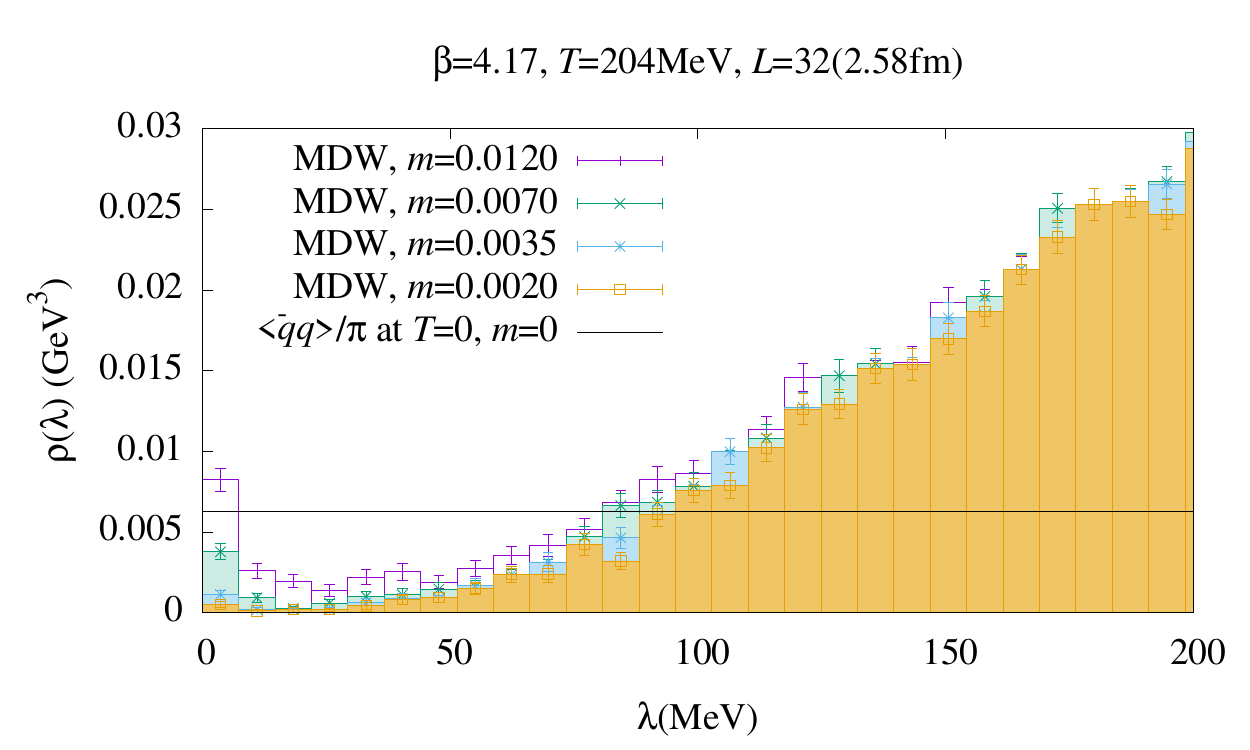}
    \end{minipage}%
    \begin{minipage}[t]{0.5\columnwidth}
    \includegraphics[clip,width=1.0\columnwidth]{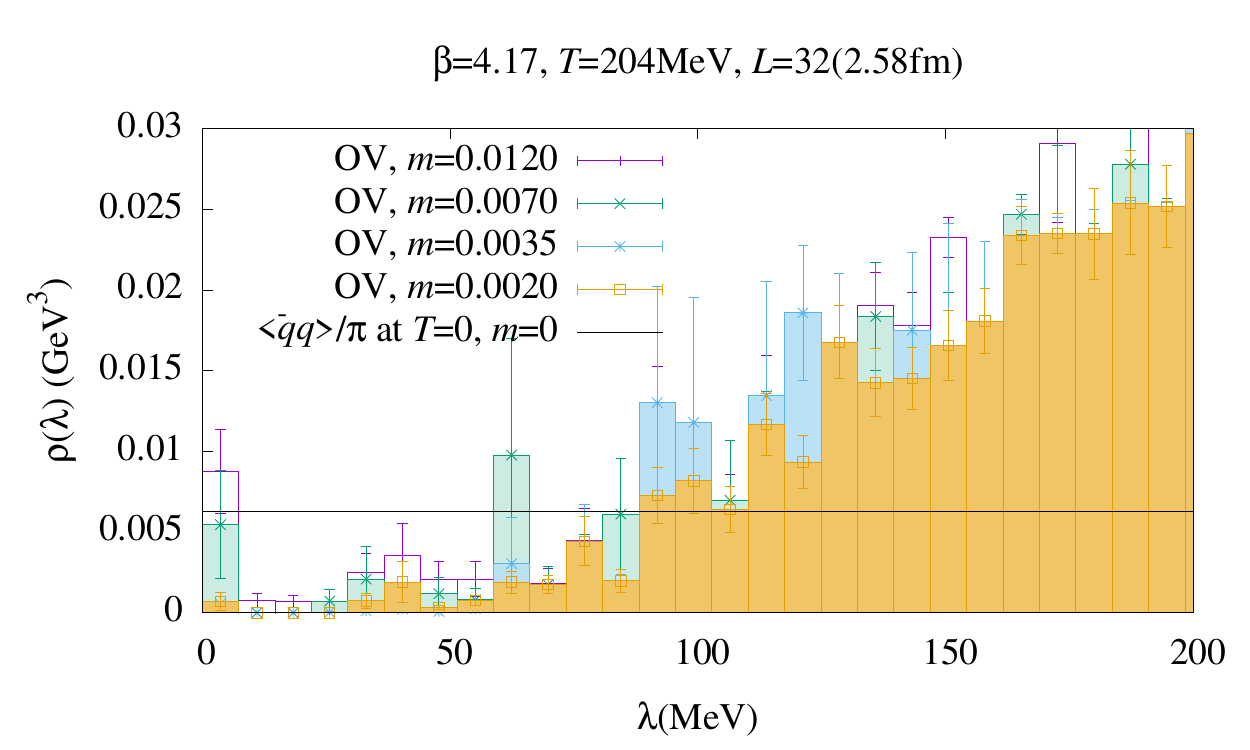}
    \end{minipage}
    \begin{minipage}[t]{0.5\columnwidth}
    \includegraphics[clip,width=1.0\columnwidth]{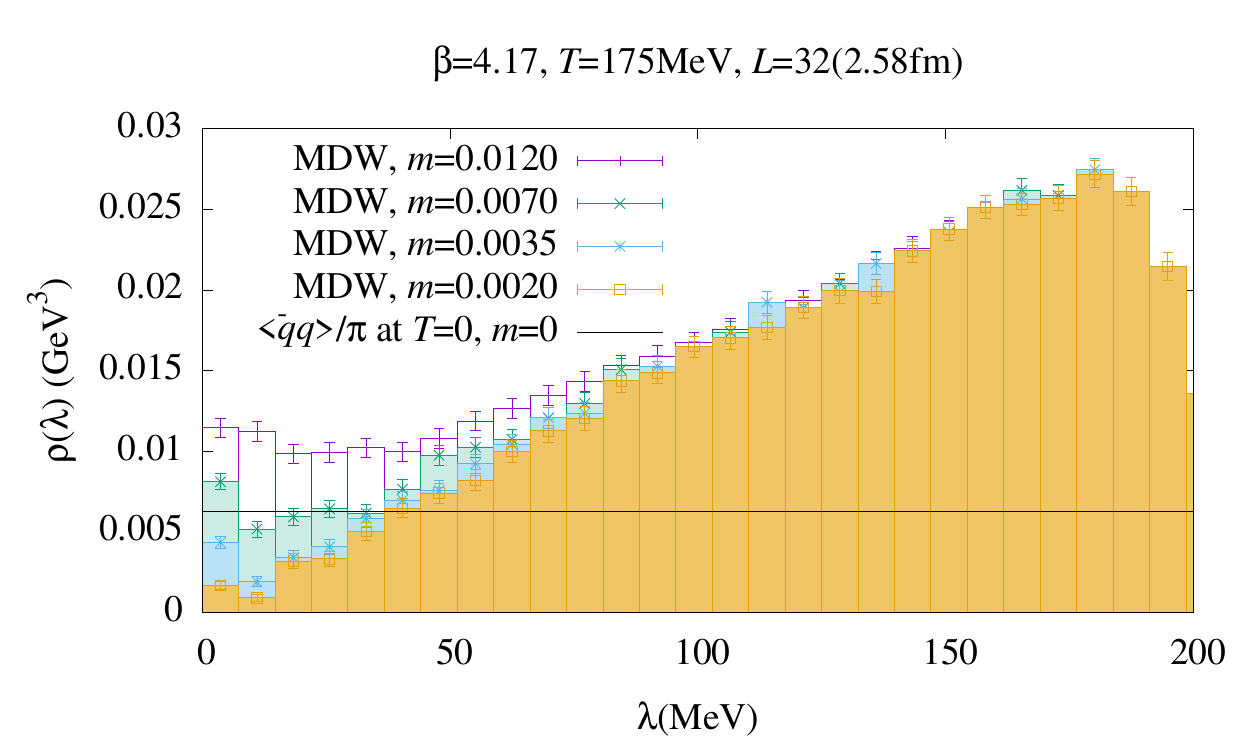}
    \end{minipage}%
    \begin{minipage}[t]{0.5\columnwidth}
    \includegraphics[clip,width=1.0\columnwidth]{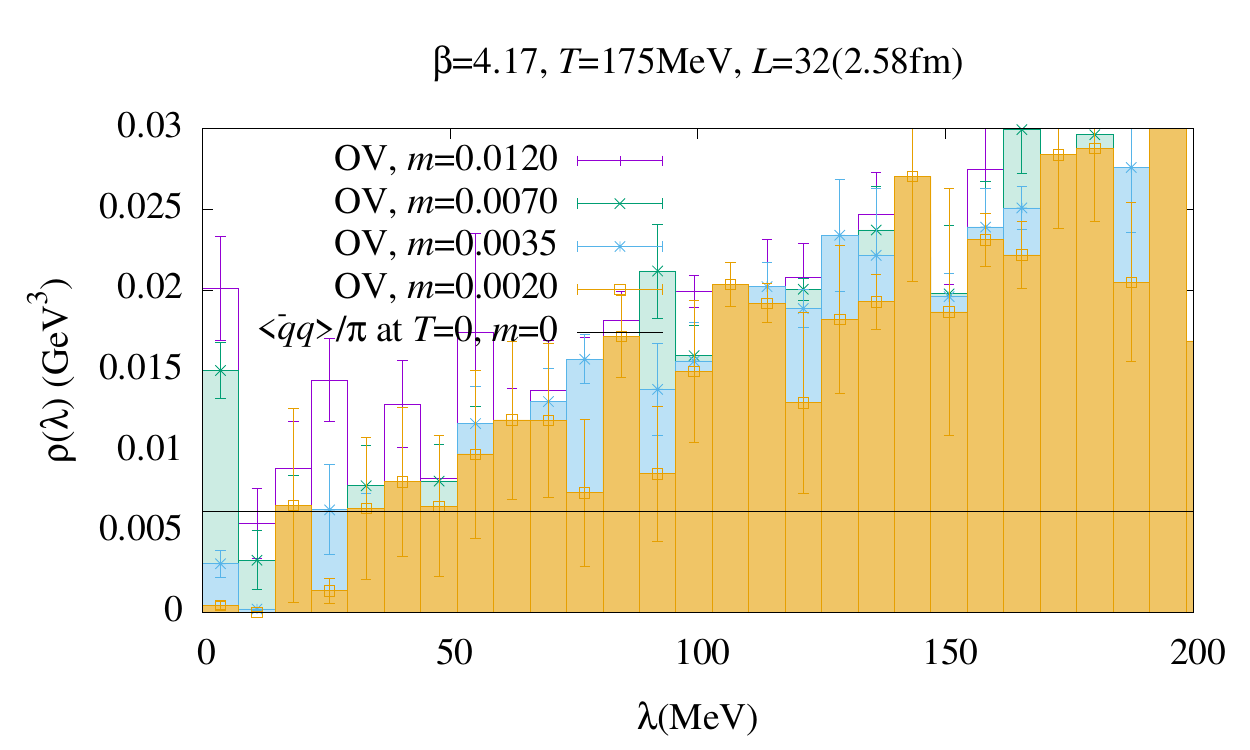}
    \end{minipage}
    \caption{Dirac spectrum $\rho(\lambda)$ at $T=204$ (upper) and $175$ MeV (lower).
    Left: the results on MDW ensembles.
    Right: the results on reweighted OV ensembles.
}
    \label{fig:spectrum}
\end{figure}

\begin{figure}[t!]
    \begin{center}
            \includegraphics[clip,width=0.5\columnwidth]{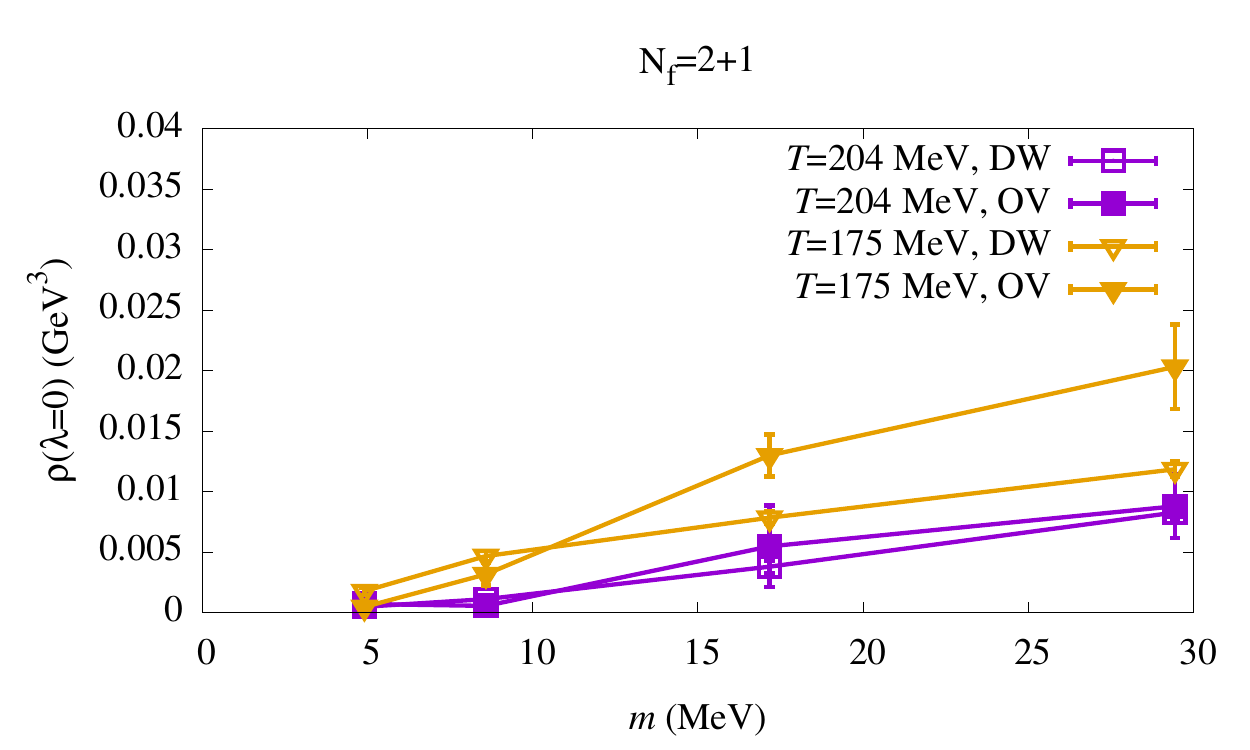}
    \end{center}
    \caption{The lowest-bin value of Dirac spectrum, $\rho(\lambda=0)$, at $T=204$ and $175$ MeV.
    Open symbols: the results on MDW ensembles.
    Filled symbols: the results on reweighted OV ensembles.
}
    \label{fig:lowest_bin}
\end{figure}

In Fig.~\ref{fig:spectrum}, we plot the spectral density of the M\"obius domain-wall
Dirac operator (in the left panels)
and that of overlap Dirac operator (right panels) at $T=204$ MeV (top
panels) and 175 MeV (bottom panels).
As a reference, we draw a horizontal line at the chiral condensate
divided by $\pi$ obtained at zero temperature in our previous work \cite{Cossu:2016eqs}.
We can see a strong suppression near $\lambda=0$ as the up-down quark
mass decreases.
In Fig.~\ref{fig:lowest_bin}, we focus on the lowest bin of the Dirac specturm and plot the data as a function of the quark mass.
We can confirm that the near-zero peak, which may be reflecting the instanton-like excitations through the chiral zero modes, is also strongly suppressed near the chiral limit.
These observation suggests that the both of $SU(2)_L\times SU(2)_R$ and $U(1)_A$ breakings are suppressed at these temperatures.

\section{$U(1)_A$ susceptibility}\label{sec-3}

Next let us discuss the $U(1)_A$ susceptibility, defined by the
difference between the two-point mesonic functions of the isovector-pseudoscalar $\pi^a$ and
isovector-scalar $\delta^a$
\begin{equation}
\Delta_{\pi-\delta} \equiv \chi_\pi - \chi_\delta \equiv \int d^4x \langle \pi^a(x) \pi^a(0) - \delta^a (x) \delta^a(0) \rangle, \label{eq:Delta_def}
\end{equation}
where the subscript $a$ denotes the isospin index.

In lattice gauge theory, we can express this quantity using the spectral decomposition of the overlap Dirac operator by
\begin{equation}
\Delta_{\pi-\delta}^{\mathrm{ov}} =  \frac{1}{V(1-m^2)^2} \left< \sum_i \frac{2m^2(1-\lambda_i^{(\mathrm{ov},m)2})^2}{\lambda_i^{(\mathrm{ov},m)4}} \right> , \label{eq:Delta_ov}
\end{equation}
where $\lambda_i^{(\mathrm{ov},m)}$ is the eigenvalue of the massive overlap Dirac operator $H_m=\gamma_5((1-m)D_\mathrm{ov}+m)$.
Here and in the following, the lattice spacing is set to unity $a=1$.
In order to cancel a possible logarithmic divergence, we employ a simple subtraction scheme
using different valence masses proposed in Refs.~\cite{Tomiya:2016jwr,Aoki:2020noz}.

\begin{figure}[t!]
    \begin{center}
           \includegraphics[clip,width=0.5\columnwidth]{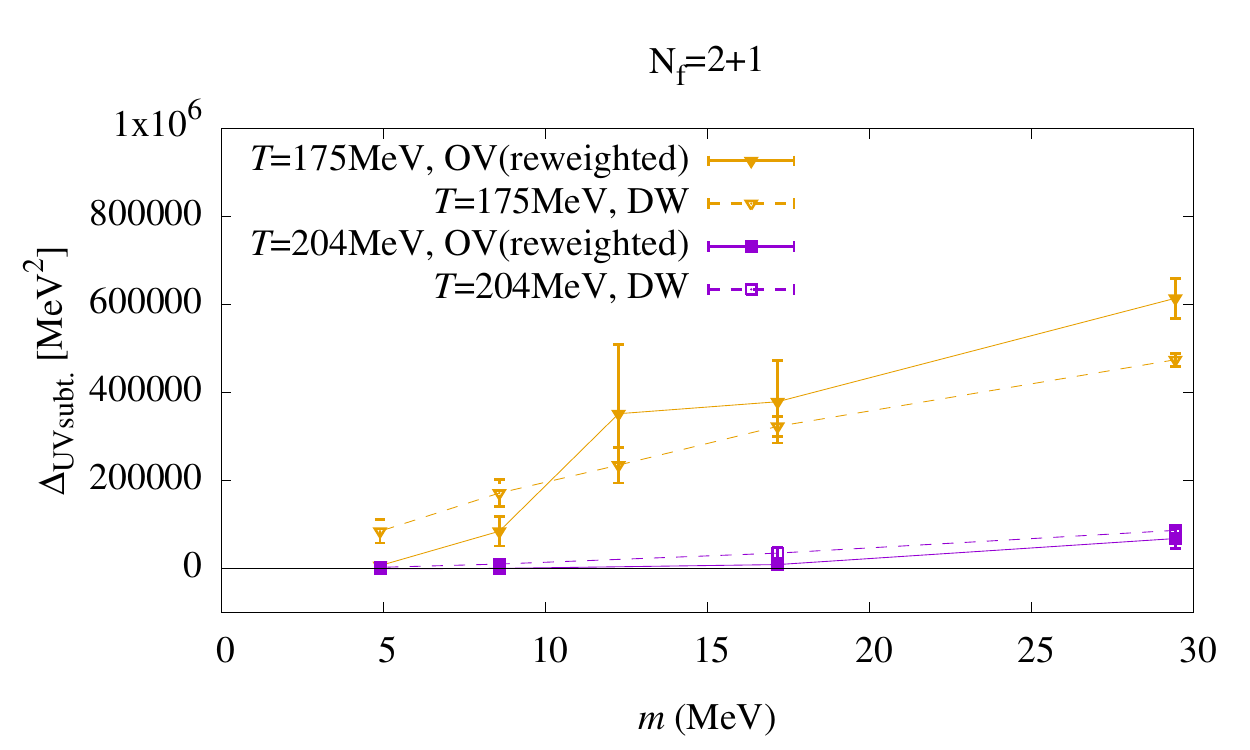}
    \end{center}
    \caption{$U(1)_A$ susceptibility $\Delta_{\pi-\delta}^{\mathrm{ov}}$ at $T=204$ and $175$ MeV.
}
    \label{fig:u1_sus}
\end{figure}

In Fig.~\ref{fig:u1_sus}, we show the quark mass dependence of  the $U(1)_A$ susceptibility at $T=204$ and $175$ MeV.
At the heaviest quark mass, we find that $\Delta_{\pi-\delta}^{\mathrm{ov}}$ is a nonzero value at both the temperatures.
As the up-down quark mass decreases, $\Delta_{\pi-\delta}^{\mathrm{ov}}$ becomes smaller to a few MeV level at the physical point $m \sim 5$ MeV.
We conclude that the axial $U(1)$ breaking is suppressed near the chiral limit, at a similar rate of $N_f=2$ QCD~\cite{Aoki:2020noz}.

\section{Topological susceptibility}\label{sec-5}

The topological susceptibility defined by
\begin{equation}
\chi_t=\frac{\langle Q_t^2 \rangle}{V}, \label{chit}
\end{equation}
where $Q_t$ is the topological charge of the gluonic fields,
is another probe of the $U(1)_A$ anomaly.
In this work, we try two types of definitions:
1) the definition by the index of the overlap Dirac operator,
\begin{equation}
Q_t=n_+ - n_-,  \label{Q_fermion}
\end{equation} 
where $n_\pm$ is the number of the zero modes with positive or negative chirality,
and 2) the geometric definition given by
\begin{equation}
Q_t (t)=\frac{1}{32\pi^2} \sum_x \varepsilon^{\mu \nu \rho \sigma} \mathrm{Tr} \, \left[ F_{\mu \nu}(x,t) F_{\rho \sigma} (x,t) \right], \label{Q_gluon}
\end{equation}
with a clover-type discretization of the gluon field strength tensor $F_{\mu \nu}(x,t)$ at
spacetime $x$.
Before the measurement of 2), we perform $ta^2=5$ steps of the Wilson flow~\cite{Bruno:2014ova}.

\begin{figure}[t!]
    \begin{center}
            \includegraphics[clip,width=0.5\columnwidth]{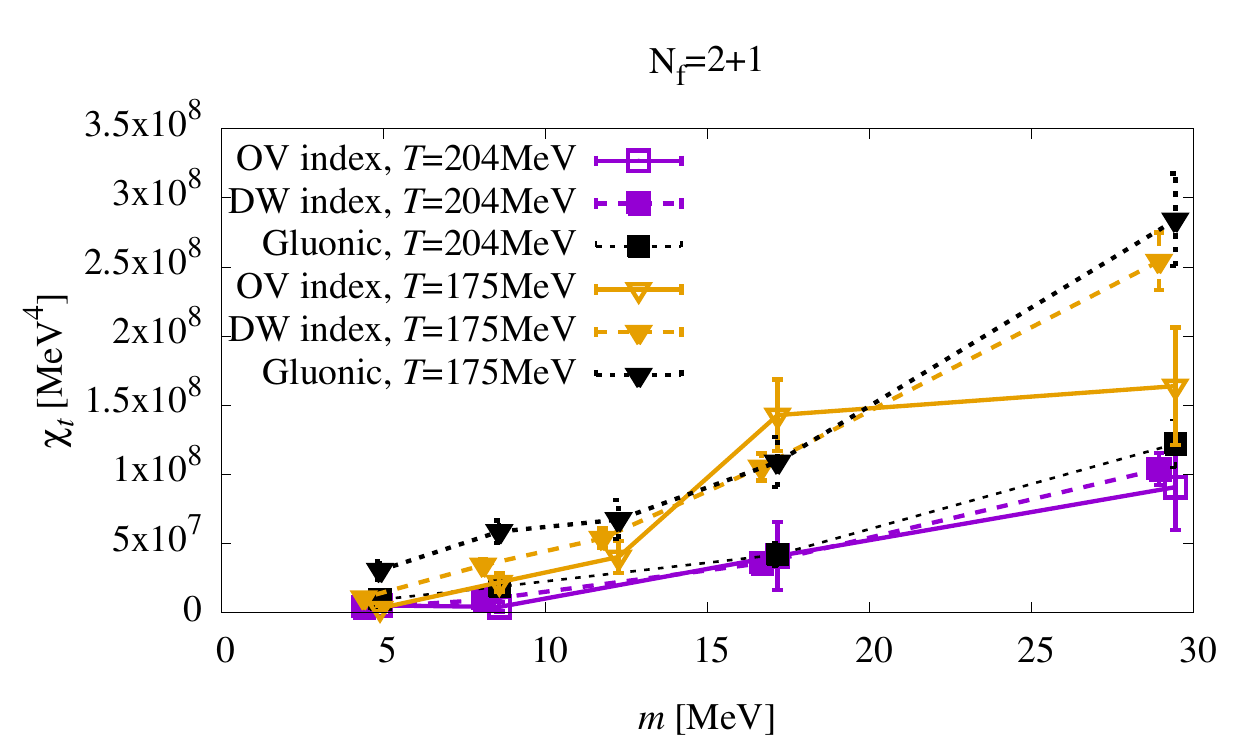}
    \end{center}
    \caption{Topological susceptibilities $\chi_t$ at $T=204$ and $175$ MeV.
    Colored points: $\chi_t$ from the fermionic definition (\ref{Q_fermion}) on MDW and reweighted OV ensembles.
    Black points: $\chi_t$ from the gluonic definition (\ref{Q_gluon}) on MDW ensembles.
}
    \label{fig:chit}
\end{figure}

In Fig.~\ref{fig:chit}, we show the results for the three types of the topological susceptibilities: fermionic
definitions on MDW and OV ensembles and the gluonic definition.
Three results are almost consistent with each other, and we find that $\chi_t$ is strongly suppressed at the lightest quark mass.

\section{Conclusion}\label{sec-6}

In these proceedings, we reported on our preliminary study of the Dirac spectra, $U(1)_A$ susceptibility, and topological susceptibility in $N_f=2+1$ lattice QCD  with dynamical chiral fermions.
At two different simulated temperatures above the critical temperature, our numerical data covering the physical point of the up-down and strange quark masses show a strong suppression of the axial $U(1)$ breaking near the chiral limit.
So far the results are obtained with a fixed lattice size $L=32$ and a fixed lattice spacing $a^{-1}=2.453$ GeV (see Ref.~\cite{Aoki:2021kbh} for a related simulation along the line of constant physics) and limited configuration numbers, but our numerical results look already good enough to show a consistency with those obtained in our $N_f=2$ QCD studies~\cite{Aoki:2020noz}.

We plan to extend this work to a temperature $T=153$ MeV, which is
near the pseudo-critical temperature of QCD at the physical point, measuring the
mesonic/baryonic observables, including a check of a further extension of the (emergent) symmetries~\cite{Glozman:2014mka, Glozman:2015qva,Rohrhofer:2017grg,Rohrhofer:2019qwq,Rohrhofer:2019qal}.
It will be also important to investigate the contribution of the axial $U(1)$ breaking to the quark mass dependence of the chiral condensate or chiral
susceptibility~\cite{Aoki:2021qws, Aoki:2021nur}.

\section*{Acknowledgment}\label{sec-Ack}
Numerical simulations are performed on IBM System Blue Gene Solution at KEK under a support of its Large Scale Simulation Program (No. 16/17-14) and Oakforest-PACS at JCAHPC under a support of the HPCI System Research Projects (Project IDs: hp170061, hp180061, hp190090, and hp200086) and Multidisciplinary Cooperative Research Program in CCS, University of Tsukuba (Project IDs: xg17i032 and xg18i023).
This work is supported in part by the Japanese Grant-in-Aid for Scientific Research (No. JP26247043, JP18H01216 and JP18H04484), and by MEXT as ``Priority Issue on Post-K computer" (Elucidation of the Fundamental Laws and Evolution of the Universe) and by Joint Institute for Computational Fundamental Science (JICFuS).

\bibliographystyle{JHEP}
\bibliography{lattice2021}
\end{document}